# HOW TO COMMISSION, OPERATE AND MAINTAIN A LARGE FUTURE ACCELERATOR COMPLEX FROM FAR REMOTE SITES

P. Czarapata FNAL, D. Hartill, Cornell; S. Myers, CERN; S. Peggs, BNL; N. Phinney, SLAC; N. Toge, KEK; F. Willeke, DESY; C. Zhang , IHEP Beijing

*Abstract*

A study on future large accelerators [1] has considered a facility, which is designed, built and operated by a worldwide collaboration of equal partner institutions, and which is remote from most of these institutions. All operation modes were considered including trouble shooting, development, commissioning, maintenance, and repair. Experience from existing accelerators shows that most of these activities are already performed 'remotely'. The large high-energy physics and astronomy experiments already involve international collaborations of distant institutions. Based on this experience, the prospects for a machine operated remotely from far sites are encouraging. Experts from each laboratory would remain at their home institution but continue to participate in the operation of the machine after construction. Experts are required to be on site only during initial commissioning and for particularly difficult problems. Repairs require an on-site non-expert maintenance crew. Most of the interventions can be made without an expert and many of the rest resolved with remote assistance. There appears to be no technical obstacle to controlling an accelerator from a distance. The major challenge is to solve the complex management and communication problems.

## 1 INTRODUCTION

The next generation of particle accelerators may require a new mode of international and inter-laboratory collaboration since they are too costly to be funded by a single nation and too large to be built by a single laboratory. The tremendous technical challenge of a new facility requires a critical mass of highly qualified and experienced physicists and engineers. These experts are presently distributed among the accelerator centers around the world and it is believed important to maintain and develop this broad base of expertise. The successful recent accelerator technology development depended on extensive exchange of people with complementary technical skills. Therefore, it is desirable that several accelerator laboratories will participate in any future project. A consequence of a multi-laboratory project is that the facility will be located a considerable distance from most of the contributing institutions which design and build it. Shared remote operation is a model that allows the experts who designed and built the machine to continue to participate in its operation. In order to make such a model work, the collaborating institutions must have a continuing commitment to the project. We discuss below a model for an international multi-laboratory collaboration to construct and operate an accelerator facility, which attempts to meet this requirement. The issues for far-remote operation are based on this model. The following questions are addressed: What is required for effective exchange of experience, ideas, parameters and data necessary for adequate discussion of the problems expected during commissioning, tune-up, failure analysis, and performance and reliability improvements? What local staff is required for operations, maintenance, and repair? What needs to be changed in the technical design of the hardware components to allow remote diagnosis and analysis? Are the costs of these changes significant? What are the requirements on the control system data transmission speed or bandwidth to support remote operation? Are presently available technologies a limitation requiring further research and development?

## 2 AVAILABLE EXPERIENCE

Existing large accelerators such as LEP and HERA are remotely operated where the controls architecture supports 'far-remote' control. Feedback loops that require fast response are implemented locally and do not require continuous intervention from the main control room. Analog signals are almost always digitized before transmission to the control room so that there is no loss of information through long cable runs. The enormous advances in computing and networking have made digital media the most convenient and inexpensive method for transmitting data, even over short distances. The large size of present accelerators and the limited access demands that interventions be well-planned and undertaken only after extensive remote diagnostics. Non-expert maintenance staff is able to handle most of failures and repairs. In difficult cases, they are assisted by experts via telephone or via remote computer access to the components. Unscheduled presence of experts on site is exceptional. Detailed reports and analysis from LEP and HERA that support these conclusions are available [1]. The commissioning, operation and optimization of the SLC is perhaps the most relevant experience for a future linear collider. However, because the SLC was an upgrade of the existing SLAC linac, many of the technical components were not modern enough to support remote maintenance and troubleshooting. A significant presence of expert staff on site was required. The control system was designed to allow consoles to be run remotely from home or office. With proper coordination, they could be run from other laboratories. However, operators in the SLC control center relied on many analog signals from older diagnostics, which were not available remotely.

Although extensive feedback systems were developed for the SLC to stabilize the beam parameters and optimize luminosity, some tasks still required frequent operator. This experience might seem discouraging for the feasibility of far-remote operations, but none of these technical limitations are fundamental given modern technology. The increase in SLC performance was often enabled by a good data logging systems and off line analysis. Such analysis could have been performed from anywhere in the world if the data were available to an external group. In fact, many aspects of SLC experience with feedback, automated procedures and complex analysis are encouraging for far-remote operation.

Many of the initial difficulties in commissioning accelerators are caused by insufficient diagnostics. Whenever comprehensive controls and diagnostics have been available in the control room at an early stage of accelerator commissioning, they have facilitated a rather smooth and quick turn-on, as seen at ESRF and PEP II. There are many more examples of facilities with insufficient initial diagnostics where progress was unsatisfactory. Any large accelerator must have remote troubleshooting capability, simply because of the distances involved. The conclusion is that any facility with adequate diagnostics and controls for efficient operation could also easily be commissioned remotely.

Non-accelerator projects also have extensive experience with remote operation of complex technical systems with restricted accessibility. The successful operation of space experiments and of distant telescopes demonstrates that efficient remote operation is possible, practicable and routinely performed. In particular, many observatories are built in rather inhospitable locations and operated with only very little technical support on site. Troubleshooting and consultation with experts is almost exclusively performed remotely. The European space agency ESO has remotely operated telescopes in Chile from a control center in Germany for a decade. Their operational experience [2] is encouraging but demonstrates that the main difficulties lie in handling exceptional situations. These institutions maintain a strong presence of experts on site despite the unfavorable conditions in order to mitigate these problems. The collaborators on a remote accelerator project should carefully analyze and learn from the ESO experience.

# 3 MODEL OF REMOTE OPERATION

## 3.1 General Organization

The accelerator is built and operated by a consortium of institutes, laboratories or groups of laboratories. Each collaborator is responsible for a complete section of the machine including all subsystems. This responsibility includes design, construction, testing, commissioning, participation in operations, planning and execution of machine development, maintenance, diagnosis and repair of faulty components. These machine sections are large contiguous parts of the machine such as for example injectors, damping rings, main linacs, beam delivery, for a linear collider. This eases the design, construction, and operation of the accelerator, if the responsibility for large systems is assumed by a group of institutions from one region. To minimize the variety of hardware types to be operated and maintained, collaborators are also responsible for a particular category of hardware spanning several geographic regions.

Central management is needed to coordinate the design and construction of the machine and later supervise operation and maintenance. Responsibilities include the overall layout of the accelerator with a consistent set of parameters and performance goals. This group ensures that all components of the accelerator fit together and comply with the requirements for high performance and efficient operation (definition of naming conventions, hardware standards, reliability requirements, quality control, control system standards and interfaces as well as of the real-time and off line database). The central management coordinates the construction schedule and has responsibility for the common infrastructure (roads, buildings and tunnels, power, water distribution, heating and air conditioning, cryogenics, miscellaneous supplies, site-wide communications and networks, radiation and general safety).

Central management also plans and coordinates the commissioning, as well as supervision and training of local maintenance crews. A central operation board (in coordination with the experiments) would be responsible for the mode of operation, operational parameters, machine study periods, and interventions, planning of maintenance periods, organization of machine operation, and training of the operations crews. All remote operation crews would report to the central operation board. Regardless of where the active control center is located, high performance operation of the accelerator will depend on a continuous flow of information and input from all of the collaborators. They must maintain responsibility for the performance of their components for the entire operational lifetime of the machine.

## 3.2 Machine Operation

In the multi-laboratory model, there are several fully functioning control centers capable of operating the entire accelerator complex, one at the site and one at each of the collaborating institutions. The operations crew is decentralized and can operate the accelerator from a far-remote center. At any given time, however, the machine is operated from only one of these control centers. The current control center has responsibility for all aspects of accelerator operation including commissioning, routine operation for physics, machine development studies, ongoing diagnosis, and coordination of maintenance, repairs and interventions. Control is handed off between centers at whatever intervals are found to be operationally effective. Supporting activities may take place at the other locations if authorized by the active control center.

## 3.3 Maintenance

The collaborators remain responsible for the components they have built. They must provide an on-call service for remote troubleshooting. The current operations crew works with the appropriate experts at their home institutions to diagnose problems. It has the authority to respond to failures requiring immediate attention. An on-site crew is responsible for exchanging and handling failed components. Their responsibilities include putting components safely out of operation, small repairs, disassembling a faulty component or module and replacing it by a spare, assisting the remote engineer with diagnosis, shipment of failed components to the responsible institution for repair, maintenance of a spares inventory, putting the component back into operation and releasing the component for remotely controlled turn-on and setup procedures. Some tasks such as vacuum system interventions or klystron replacement will require specialized maintenance staff which must be available on site to provide rapid response time. Decisions about planned interventions must be made by the operations board in close collaboration with the laboratory responsible for the particular part of the machine.

### 3.4 Radiation and other Safety Issues

Any accelerator can produce ionizing radiation. Its operation must be under strict control. In addition to the laws and requirements of the host country and its overseeing government agencies, there are also the internal rules of the partner laboratories. Usually it is required that on-site staff be supervising beam operation to guarantee responsibility and accountability and permit safe access to the accelerator housing. There are also concerns about the activation of high-power electrical devices and other potentially hazardous systems requiring interlocks and tight control. We believe that there exist straightforward technical solutions to ensure the required safety and security. The legal and regulatory issues are more difficult and will need careful investigation. Most likely, a near-by laboratory will have to assume responsibility for radiation and other safety issues. Similarly, unusual events like fires, floods, accidents, breakdown of vital supplies or general catastrophes will require a local crisis management team available on call to provide an effective on-site response. There must be a formal procedure to transfer responsibility to the local crisis management in such instances. This function could be provided by nearby collaborating institutions.

## 4 REQUIREMENTS FOR FAR-REMOTE OPERATION

### 4.1 Organizational Requirements

Operation via a rotating control center requires good documentation and a mechanism to assure continuity when the operations are handed over to another laboratory. Electronic logbooks will be necessary, including a comprehensive log of all commands and events with time stamps, information on the originator and comments, with a powerful intelligent browser to make the logged information useable for analysis. All control rooms should be close to identical with a no local dialect and specialization. A comprehensive system of tokens or permissions to coordinate between the different control centers is needed. These are granted by the operators in charge. The tokens must have sufficient granularity to allow active remote access to a specific component and to regulate the level of intervention. Some organizational measures will have to be taken to avoid misunderstanding and loss of operation time. This includes a more formal use of language with strictly and unambiguously defined elements. Formal names are required for accelerator sections, lattice elements, technical components and even buildings. This means that a comprehensive dictionary has to be written and maintained.

In order to keep all collaborators well informed and involved and in order to maintain active participation of distant institutions, a special effort should be made to make the control center activities visible and transparent for distant observers, in particular the current status of machine operations. Monitors should be available to follow the operations progress and discussions in the active control center. They should easily allow regular 'visits' to the control center. All operations meetings (shift change, ad hoc meetings for troubleshooting, operation summaries, coordination with experiments, etc.) should be open to remote collaborators. Virtual 'face-to-face' communications can support multi-party conversations, including shared 'blackboards' and computer windows, perhaps using virtual 'rooms' to accommodate specialists with common interests. A permanent videoconference type of meeting may serve as a model. We expect that growing commercial interest in this sector will promote the needed development.

To avoid having the accelerator teams cluster in local groups with limited exchange of information, one should prepare for regular exchange of personnel, for plenary collaboration meetings, for common operator training across the collaboration and for regular exchange and rotation of leadership roles in the project.

A relatively modest staff appears to be required on site for operations, maintenance or repair. Most of the activities of operation, troubleshooting and failure diagnosis can be performed remotely by off-site personnel. Extrapolating from the experience of existing facilities, expert intervention on site is required in only about 1% of the failures. If one assumes a rate of 2000 incidents per year, there should be not more than 20 occasions where expert help has to be available on site even without relying on further improvements in remote diagnostics, increased modularity and more maintenance friendly future designs. Extrapolating from HERA experience, the size of the on-site maintenance crew is estimated to be about 75 persons for a large future facility. For efficient operation of the accelerator, regular maintenance is required in addition to the repair of failed components. This work must also be performed by on-site staff that is supported by a nearby laboratory or by industrial contractors. The collaborator responsible for the components would plan and manage these efforts under the coordination of the operation board. A small coordination team of about ten would be needed to provide the necessary micromanagement. In addition, there must be staff on site for security, for radiation safety, and for maintenance of infrastructure, buildings and roads. The number of persons needed depends very much on the specific circumstances of the site and the type of accelerator and it is hard to predict a number. In a large laboratory, the staff for these tasks is typically 50-100. In conclusion, we estimate that a local staff of about 200 would be required to maintain the facility and assure operations.

## 4.2 Technical Requirements

The control system must optimize the flow of information from the hardware to the operations consoles to provide remote accessibility of the hardware from remote sites without excessive data rates. The layered approach of modern control systems comfortably supports these requirements.

The console applications at the control centers would essentially only build up displays and relay operator commands. These activities require a slower data rate commensurate with human response times, which should not be a problem over any distance on earth. The requirements for console support are well within the reach of existing technology. The most significant bandwidth demand is for real-time signals, which are used for continuous monitoring by the operations staff. Most of the existing accelerator control systems use Ethernet LAN technology for data communications at the console level. In present facilities, 10Mbit/sec Ethernet technology is sufficient to accommodate the required data rate with an overhead of a factor of ten. The technology for ten times this bandwidth is already available and further development can be anticipated. This should be more than adequate for any future console communication requirements.

The intercontinental data connections have been revolutionized by the recent progress in fiber optics systems providing data rates in the multi-Tbit/sec range, or nearly inexhaustible capabilities. Future needs for data communications at the particle laboratories are in the range of several Gbit/sec [3]. They are driven by the exchange of experimental data. The need for remote accelerator control is in the order of a few 10Mbit/sec which doesn't constitute a significant fraction of the anticipated connectivity. Thus the network is not expected to impose any limitation to remote operations.

High performance accelerators rely extensively on automated procedures and closed loop control. These functions often require high speed or high bandwidth and therefore would all be implemented in the on-site layers of the control system, as would time-critical machine protection algorithms, extensive data logging and execution of routine procedures.

The evolution of computer hardware and networks has allowed a migration of computing power from large centralized systems to highly distributed systems. This evolution matches well the growing accelerator complexes. Networks with Gigabit speeds and processors with clock speeds approaching one GHz have pushed far greater control autonomy to lower levels in the controls architecture. These developments favor a 'flat' (non-hierarchical) network structure with intelligent devices that would be directly accessible over the network. Such devices essentially coincide with the current catchword 'Network Appliance', and there will be an immense amount of commercial activity in this direction which will be useful for future projects.

The intelligent device model also implies that the devices be directly on the network rather than hanging on a field-bus below some other device. Traffic can be localized in this structure using 'switches' which forward packets only to the port on which the destination device hangs and whose 'store and forward' capability essentially eliminates Ethernet collisions.

On the accelerator site, the majority of repairs would involve the exchange of modules. This requires that all components be composed of modules of a reasonable, transportable size which have relatively easy to restore interfaces to the other constituents of the component.

On the other hand, the requirements for the hardware components of a remotely operated accelerator are essentially identical to the requirements for any large complex technical facility. The general design criteria are: redundancy of critical parts, if cost considerations allow it; avoidance of single point failures and comprehensive failure analysis; over-engineering of critical components to enhance mean time between failure; standardization of design procedures; quality assurance testing; documentation; standardization of components, parts and material wherever technically reasonable; avoidance of large temperature gradients and thermal stress; and control of humidity and environmental temperature extremes.

Specific features connected to remote operation are foreseen: high modularity of the components to ease troubleshooting and minimize repair time; more complete remote diagnostics with access to all critical test and measurement points necessary to reliably diagnose any failure; and provision for simultaneous operation and observation. If a device is to be fully diagnosable remotely, it is important that a detailed analysis of the requirements be an integral part of the conceptual design of the component.

A survey of engineers and designers in the major accelerator laboratories indicates that all of these design goals are already incorporated in planning for future accelerators. Due to the large number of components, even with an extremely high mean time between failures, one must expect several breakdown events per day. Even for an accelerator that is integrated into an existing laboratory, comprehensive remote diagnostics are obviously necessary to minimize downtime. This will be one of the crucial technical issues for a large new facility. The mean time between failures has to improve by a factor of 5-10 compared to existing facilities like HERA. This is the real challenge and any additional requirements for remote operation are minor by comparison.

The conclusion is that the major technical challenges for the hardware of a future accelerator are due to the large number of components and the required reliability and not to the possibility of remote operation and diagnostics. The additional costs for compatibility with remote operation appear negligible.

## 5. SUMMARY AND CONCLUSION

We consider a facility which is remote from most of the collaborating institutions, designed, built and operated by a worldwide collaboration of equal partner institutions. Expert staff from each laboratory remains based at its home institution but continues to participate in the operation of the machine after construction. We consider all operation activities. As far as maintenance, troubleshooting and repair is concerned, the experience from existing laboratories is encouraging, indicating that most activities are already performed 'remotely', or could be with properly designed equipment. The experts are only rarely required to be physically present on site. Repairs require an on-site maintenance crew. Most of the interventions are made without an expert or with only telephone assistance. For a future large accelerator facility, we conclude that it should be possible to perform most of the tasks remotely. Maintenance, troubleshooting and repair by non-experts do require comprehensive remote diagnostics, modular design of components, and a high level of standardization. An accelerator could be operated far-remotely. Modern control systems use a layered approach, which appears to be adequate. The rapid rate of development of communications technology should easily support the demands of future accelerator operation. Considering this we conclude that there appears to be no technical obstacle to far-remote control of an accelerator.

Operation of the accelerator is not an easy task. Frontier facilities are inevitably pushing the limits of accelerator technology and present unanticipated difficulties that require intense effort from a dedicated team of experts to diagnose and solve each new problem. Past experience has shown how critical it is for these experts to have offices near each other to facilitate exchange of ideas and information. Equally important is contact between the experimenters and the accelerator physicists, and between the physicists, engineers and operations staff. To encourage an effective interchange between these disparate groups, it will be necessary to have a critical mass of experts located in at least one, if not several, of the laboratories.

During normal operation, the on-site staff required could be much smaller than are typically at existing large laboratories. A reliable number for the minimum staff depends very much on the details of the remote facility but experience from large machines indicates that is could be as small as 200. There would be a much greater number of technical staff of all descriptions actively involved in the accelerator operation remotely.

The major challenge of a remote operation model lies in solving the complex management and communication problems.

## ACKNOWLEDGEMENTS

The authors acknowledge that much of the material presented in this report was supplied by Prof. M. Huber, ESA; Dr. M. Ziebell, ESO, M. Jablonka and B. Aune, CEA Saclay, G. Mazitelli, INFN, C. Pirotte, CERN, S. Herb and M. Clausen, H. Freese, M. Bieler, J.P. Jensen, M. Staak, F. Mittag, W. Habbe, M. Nagl, and J. Eckolt, DESY.